\documentstyle[11pt,newpasp,twoside,psfig]{article}
\def\edcomment#1{\iffalse\marginpar{\raggedright\sl#1\/}\else\relax\fi}
\reversemarginpar

\begin{document}
\title{Self-consistent ab initio calculations for photoionization and
electron-ion recombination using the R-matrix method}
\author{Sultana N. Nahar}
\affil{Dept of Astronomy. Ohio State University, Columbus, Ohio 43210,
USA }

\begin{abstract}
Most astrophysical plasmas entail a balance between ionization and
recombination. We present new results from the powerful R-matrix method
which yields in an ab initio manner: (I) self-consistent photoionization
and recombination cross sections using identical wavefunction expansions,
(II) unified e-ion recombination rates at
all temperatures of interest, incorporating non-resonant and
resonant processes, radiative and dielectronic recombination (RR and
DR), and (III) level-specific recombination rates for many excited
atomic levels. RR and DR processes can not be measured or observed
independently. In contrast to the shortcomings of simple (but
computationally easy) approximations that unphysically treat RR and DR
separately with different methods, emphasizing marginal effects for
selected ions over small energy ranges, the R-matrix method naturally
and compeletely accounts for e-ion recombination for all atomic systems.

Photoionization and recombination cross sections are compared with
state-of-the-art experiments on synchrotron radiation sources and
ion storage rings. Overall agreement between theory and experiments is
within 10-20 \%. For photoionization the comparison includes not only the
ground state but also the metastable states, with highly resolved resonance
structures. The recent experiments therefore support the estimated accuracy
of the vast amount of photoionization data computed under the Opacity Project
(OP), the Iron Project (IP), and related works using the R-matrix method.

\end{abstract}

The inverse processes of bound-free transitions may
proceed as,

\noindent
i) Photoionization (PI) and Radiative Recombination (RR) :
$$X^{+Z} + h\nu \rightleftharpoons X^{+Z+1} + \epsilon $$
\noindent
ii) Autoionization (AI) and Dielectronic recombination (DR):
$$e + X^{+Z} \rightarrow (X^{+Z-1})^{**} \rightarrow \left\{ \begin{array}{ll}
e + X^{+Z} & \mbox{AI} \\ X^{+Z-1} + h\nu & \mbox{DR} \end{array}
\right. $$
\noindent
The doubly excited autoionizing states introduce resonances in
the cross sections. RR and DR are observed together in nature although
one may dominate the other, and quantum mechanical interference between
the two may vary from low-Z to high Z elements, also depending on ion
charge z. A new unified method has been developed (Nahar \&
Pradhan 1994, 1995; Zhang, Nahar \& Pradhan 1999) that subsumes both RR
and DR, and enables a self-consistent treatment for photoionization
and (e~+~ion) recombination in astrophysical plasmas.
 Another review article in these proceedings discusses the throretical
background of photoionization and total electron-ion recombination.
We briefly note the points germaine to the
self-consistent formulation using the R-matrix method in the close
coupling approximation: (I) identical wavefunction
expansion for the (e+ion) system is employed in ab initio
calculations for photoionization and recombination, and (II)
self-consistency {\it requires} an unified approach for (e+ion) 
recombination: $ \alpha_{R} \rightarrow \alpha_{RR} + \alpha_{DR} $,
as the natural inverse photoionization, (III) level-specific data is
specifically obtained.

 In contrast to the R-matrix formulation of the unified (e+ion)
recombination, simpler methods treat RR and DR separately in
the independent resonance approximation (e.g. Gorczyca et al. 2002).
While this unphysical separation may not be too inaccurate
for some highly charged ions, it must necessarily fail for all ions
with strong coupling, such as the low ionization stages of
iron Fe~I~-~V (see the review article by the author).
Simpler methods also tend to unduly emphasize issues of marginal
general importance, such as radiation damping of a few high-$n$
resonances, in the calculation of total recombination rates.

Fig. 1 presents photoionization cross sections, $\sigma_{PI}$, of the
ground $2s^22p^2(^3P)$ and excited $2s^22p3d(^3P^o)$ states of Fe XXI
(Nahar 2002, in preparation). The arrows show Rydberg autoionizing
resonances arising from the 8 $n$=2 levels of the core ion (8CC
wavefunction expansion). The ground state state shows significant 
resonances below the $n$=2 levels only. However, the large resonant 
features for the excited $^3P^o$ state are due to the $n$=3 thresholds, 
and enhance the effective cross section
considerably at energies much higher than the $n$=2.

\begin{figure} %
\psfig{figure=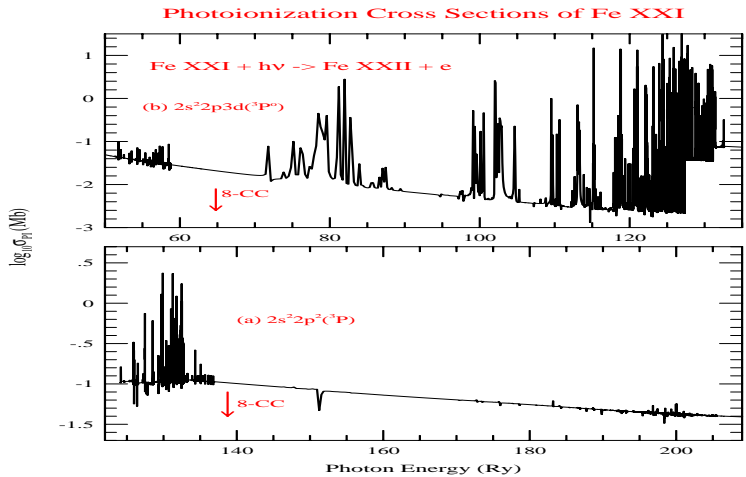,height=7.0cm,width=12.0cm}
\caption{Photoionization cross sections, $\sigma_{PI}$, of ground
$2s^22p^2(^3P)$ and excited $2s^22p3d(^3P^o)$ states of
Fe XXI (Nahar 2002).}
\end{figure}

The R-matrix results may be compared to state-of-the-art experiments 
on synchrotron radiation sources for photoionization, and on heavy ion 
storage rings for recombination, which display heretofore 
unprecedented detail in resonances and background cross sections, and 
thereby calibrate the theoretical data precisely. An example of the 
comparison between the theory and experiment for $\sigma_{PI}$ 
of ground and metastable states, $2s^22p^3(^4S^o,^2D^o,^2P^o)$, 
of O II is shown in Fig. 2 (Covington et al. 2001). Theoretical results 
(bottom panel) were obtained before the experiments were carried out, 
and helped to identify the resonances for the $^2P^o$ (green curve), 
and $^2D^o$ (blue curve) states respectively, in the total experimentally
measured cross section (top panal).

\begin{figure} %
\vspace*{-0.5cm}
\psfig{figure=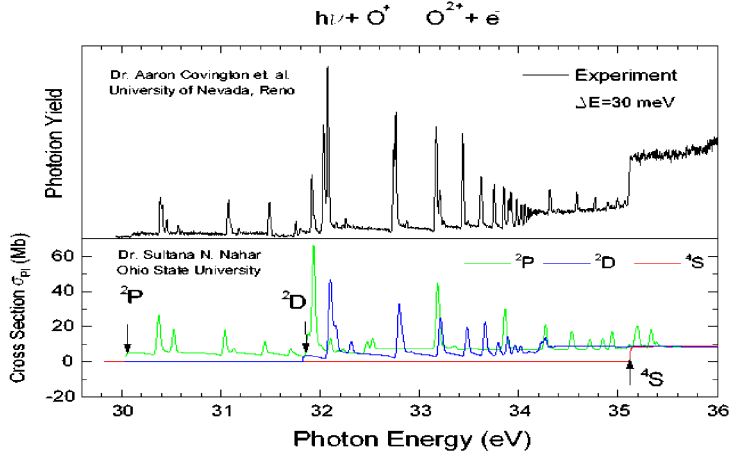,height=7.0cm,width=12.0cm}
\caption{Comparison of theoretical and experimental photoionization cross
sections of the metastable $2s^22p^3(^2P^o,^2D^o)$ states of O II.}
\end{figure}

The unified total electron-ion recombination rate coefficient, 
$\alpha_R$, is obtained from the sum of photorecombination contributions 
to a large number of bound states, and to the rest of all highly 
excited states up to $n=\infty$ via DR. Fig. 3 presents $\alpha_R$ of 
e + Ar VI $\rightarrow$ Ar V (Nahar 2000). Total unified 
-red; RR - blue; DR (high-T) - green. Typically, the total $\alpha_R$ 
is high at low T (RR limit), and decreases before rising
again due to dominance by DR. However, low-T "bumps" may
exist due to autoionizing resonances in the low-energy region.

\begin{figure} %
\vspace*{-0.3cm}
\psfig{figure=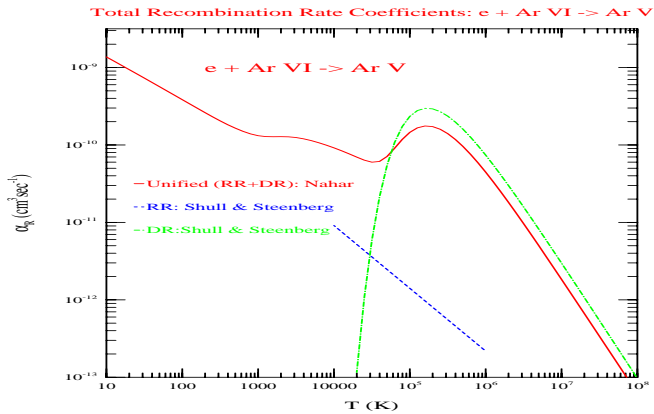,height=6.0cm,width=10.0cm}
\vspace*{-0.5cm}
\caption{Total unified recombination rate coefficient of Ar V.}
\end{figure}

The unified recombination cross sections ($\sigma_{RC}$) and rates 
have been benchmarded with experimental measurements, to about 10-15\%.  
Fig. 4(a) shows the detailed unified $\sigma_{RC}$ for (e + C IV)  
in the $1s^2 2s (^2S_{1/2} -- 1s^2 2p (^2P^o_{1/2,3/2})$ region. The 
computed rate coefficient $v \ast \sigma_{R}$ is convolved with a gaussian
experimental beamwidth (Fig. 4b), and compared with experimental results
from the heavy-ion Test Storage Ring (Fig. 4c, Schippers et al. 2001).
The present unified $\sigma_{RC}$ in 4(a) show considerably more
detail than the experimental results, but the convolved results agree
to $\sim 15$\% for individual $n$-complexes of resonances.

\begin{figure} %
\psfig{figure=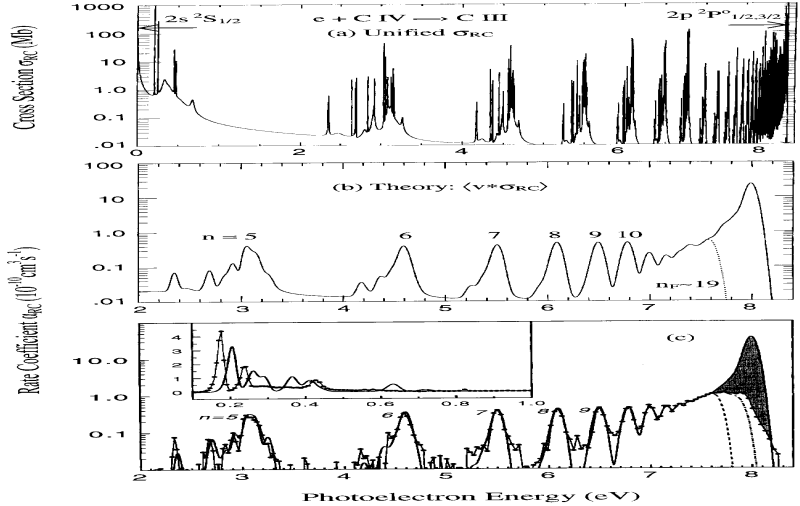,height=7.0cm,width=11.0cm}
\caption{Comparison of theoretical and experimental recombination rate
of C III (Pradhan et al. 2001).}
\end{figure}

Self-consistent sets of atomic data for photoionization, recombination 
are obtained and should lead to more accurate astrophysical 
photoionization models.

\acknowledgements{This work is supported partially by the U.S.
National Science Foundation and NASA.}

\end{document}